%% file: NaturalMssmThermalTunneling.tex
\documentclass[5p]{elsarticle}

\pdfoutput=1

\usepackage{graphicx}
\usepackage{graphics}
\usepackage[mathscr]{eucal}
\usepackage{amssymb}
\usepackage{amsmath}
\usepackage{xspace}
\usepackage{listings}
\usepackage{ulem}
\usepackage{color}
\usepackage{ulem}

\definecolor{darkgreen}{rgb}{0,0.5,0}

\biboptions{sort&compress}

\newcommand{\vev}[0]{VEV\xspace}
\newcommand{\vevs}[0]{VEVs\xspace}

\newcommand{\gev}[0]{\text{ GeV}\xspace}

\newcommand{\Eq}[1]{eq.~(#1)\xspace}

\newcommand{\Tab}[0]{table\xspace}
\newcommand{\Fig}[0]{figure\xspace}

\newcommand{\etc}[0]{\textit{etc}.\xspace}
\newcommand{\eg}[0]{\textit{e.g}.\xspace}

\newcommand{\Ref}[0]{Ref.\xspace}
\newcommand{\Refs}[0]{Refs.\xspace}
\newcommand{\Sec}[0]{section\xspace}
\def\gsim{\raise0.3ex\hbox{$\;>$\kern-0.75em\raise-1.1ex\hbox{$\sim\;$}}}
\def\lsim{\raise0.3ex\hbox{$\;<$\kern-0.75em\raise-1.1ex\hbox{$\sim\;$}}}

\newcommand{\sarah}[0]{\texttt{SARAH}\xspace}

\newcommand{\vevacious}[0]{\texttt{Vevacious}\xspace}
\newcommand{\vcs}[0]{\vevacious}
\newcommand{\homps}[0]{\texttt{\texttt{HOM4PS2}}\xspace}
\newcommand{\pyminuit}[0]{\texttt{PyMinuit}\xspace}

\newcommand{\minuit}[0]{\texttt{MINUIT}\xspace}
\newcommand{\cosmotransitions}[0]{\texttt{CosmoTransitions}\xspace}
\newcommand{\ct}[0]{\cosmotransitions}
\newcommand{\spheno}[0]{\texttt{SPheno}\xspace}
\newcommand{\softsusy}[0]{\texttt{SoftSUSY}\xspace}
\newcommand{\suspect}[0]{\texttt{SuSpect}\xspace}
\newcommand{\suseflav}[0]{\texttt{SuSeFLAV}\xspace}
\newcommand{\slha}[0]{\texttt{SLHA}\xspace}

\newcommand{\tsq}[0]{\ensuremath{\tilde{t}}\xspace}
\newcommand{\sQ}[0]{\ensuremath{{\tsq}_{L}}\xspace}
\newcommand{\sU}[0]{\ensuremath{{\tsq}_{R}}\xspace}

\newcommand{\BOL}[1]{#1}

\newcommand\ELIELout{\bgroup\markoverwith
{\textcolor{blue}{\rule[.5ex]{2pt}{0.4pt}}}\ULon}

\begin{document}
\title{Constraining the Natural MSSM through tunneling to color-breaking vacua
 at zero and non-zero temperature}

\author[Wue]{J.\ E.\ Camargo-Molina}
\ead{jose.camargo@physik.uni-wuerzburg.de}

\author[TUM]{B.\ Garbrecht}
\ead{garbrecht@tum.de}

\author[Wue,BOL_tel]{B.\ O'Leary\corref{cor1}}
\ead{ben.oleary@physik.uni-wuerzburg.de}

\author[Wue]{W.\ Porod}
\ead{porod@physik.uni-wuerzburg.de}

\author[Bonn]{F.\ Staub}
\ead{fnstaub@th.physik.uni-bonn.de}

\cortext[cor1]{Corresponding author}

\address[Bonn]{%
Bethe Center for Theoretical Physics \& Physikalisches Institut der
 Universit\"at Bonn,
 53115 Bonn,
 Germany}

\address[TUM]{%
Physik Department T70,
 Technische Universit\"at M\"unchen,
 85748 Garching,
 Germany}

\address[Wue]{%
Institut f{\"{u}}r Theoretische Physik und Astronomie,
 Universit{\"{a}}t W{\"{u}}rzburg,
 Am Hubland,
 97074 W{\"{u}}rzburg,
 Germany}

\address[BOL_tel]{+49 93131 82475}

\fntext[fn1]{%
 This article is registered under preprint number:~/hep-ph/1405.7376.}



%
\begin{abstract}


We re-evaluate the constraints on the parameter space of the minimal
 supersymmetric standard model from tunneling to charge- and/or color-breaking
 minima, taking into account thermal corrections. We pay particular attention
 to the region known as the Natural MSSM, where the masses of the scalar
 partners of the top quarks are within an order of magnitude or so of the
 electroweak scale. These constraints arise from the interaction between these
 scalar tops and the Higgs fields, which allows the possibility of parameter
 points having deep charge- and color-breaking true vacua. In addition to
 requiring that our electro-weak-symmetry-breaking, yet QCD- and
 electromagnetism-preserving vacuum has a sufficiently long lifetime at zero
 temperature, also demanding stability against thermal tunneling further
 restricts the allowed parameter space.

\end{abstract}
\maketitle

\input{introduction}
\input{method}
\input{results}
\input{discussion}

\section*{Acknowledgments}
The authors would like to thank Raghuveer Garani, Sudhir Vempati and Debtosh
 Chowdhury for discussions concerning the differences between \suseflav and
 \spheno. This work has been supported by the DFG, research training group
 GRK~1147 and project No.~PO-1337/2-1. FS is supported by the
 BMBF PT DESY Verbundprojekt 05H2013-THEORIE ``Vergleich von LHC-Daten mit
 supersymmetrischen Modellen''. BG acknowledges support from the Gottfried
 Wilhelm Leibniz programme of the DFG, from the DFG cluster of excellence
 `Origin and Structure of the Universe' and from the National Science
 Foundation under Grant No.~NSF PHY11-25915 and thanks the KITP for
 hospitality.

\bibliography{NaturalMssmThermalTunneling.bib}
\bibliographystyle{h-physrev5}

\end{document}

%% file: introduction.tex
\section{Introduction}
\label{sec:intro}

The mechanism of spontaneous symmetry breaking by the vacuum expectation value
 for a scalar field is an essential component of the standard model of
 particle physics (SM)~\cite{Glashow:1961tr, Weinberg:1967tq, Salam:1968rm},
 which has proven itself to be an accurate description of Nature all the way to
 the tera-electron-Volt scale. The discovery of the bosonic resonance at
 $125 \gev$ at the Large Hadron Collider
 (LHC)~\cite{Aad:2012tfa, Chatrchyan:2012ufa} is consistent with the Higgs
 boson of the spontaneous symmetry breaking of the SM, leading one to take the
 issue of minimizing the scalar potential seriously.

The minimal supersymmetric extension of the SM (the MSSM) has a much more
 complex scalar potential by merit of there being many more scalar fields
 (partners for each SM fermion as well as a second Higgs $SU(2)_{L}$ doublet)
 which interact with the Higgs fields. The large effect of the extra loops on
 the mass of the Higgs boson along with the non-observation of supersymmetric
 partners thus far has led to the pragmatic region of the MSSM parameter space
 known as the Natural MSSM~\cite{Ellis:1986yg, Barbieri:1987fn, Chan:1997bi}. 
 This is the region where the masses of all the partners are very large but for
 those with the largest contributions to the Higgs
 mass~\cite{Draper:2011aa, Heinemeyer:2011aa, Brummer:2012ns, Djouadi:2013vqa,%
 Arbey:2012bp, Arganda:2012qp}, which should have masses not very far above the
 electroweak scale so that there is little finely tuned cancellation between
 loop contributions to the minimization conditions, and thus is in some sense
 natural~\cite{Kitano:2006gv, Papucci:2011wy, Wymant:2012zp, Baer:2012uy}. Thus
 the \textit{stops} \tsq (scalar partners of the top quarks) should have
 TeV-scale soft supersymmetry-breaking parameters while all others are assumed
 to have very large masses. The partners of the bottom quarks and tau leptons
 could also be in the TeV-scale, but in this letter we consider only stops,
 noting that our algorithm is trivially generalizable and is already
 implemented in the public code \vcs~\cite{Camargo-Molina:2013qva}.

While the interaction between stops and the Higgs fields allow the mass of the
 Higgs boson to reach $125 \gev$ in the MSSM, it also leads to the possibility
 of the scalar potential having undesired minima apart from the desired
 symmetry-breaking (DSB) vacuum, where only the neutral components
 of the Higgs doublets get non-zero \vevs. Even though a parameter point may be
 chosen where the scalar potential has a minimum where the stops do not have
 non-zero \vevs, there is no guarantee that this is the global minimum: there
 may be deeper charge- and color-breaking (CCB) minima to which the Universe
 may tunnel~\cite{Nilles:1982dy, AlvarezGaume:1983gj, Derendinger:1983bz,%
 Claudson:1983et, Kounnas:1983td, Drees:1985ie, Gunion:1987qv, Komatsu:1988mt,%
 Langacker:1994bc, Casas:1995pd, Casas:1996de}. However, even if the DSB vacuum
 is only metastable, the parameter point is still acceptable if the expected
 tunneling time is of the order of the age of the known
 Universe~\cite{Riotto:1995am, Kusenko:1996xt, Kusenko:1996jn}. Also, given the
 convincing success of the Big Bang theory, acceptable parameter points with
 metastable DSB vacua should also have a high probability of surviving
 tunneling to the true CCB vacua through thermal fluctuations.

In \Sec~\ref{sec:method} we lay out the algorithm by which we compute
 whether a parameter point is excluded by the DSB vacuum having a very low
 probability of surviving to the present day either by a high probability of
 critical bubbles of true vacuum forming through quantum fluctuations in our
 past light-cone at zero temperature, or by such bubbles forming through
 thermal fluctuations during the period when the Universe was at sufficiently
 high temperature. In \Sec~\ref{sec:results}, we show how much of the parameter
 space is excluded by such conditions, and compare this to previous work.
 Finally we conclude in \Sec~\ref{sec:discussion}.

%% file: method.tex
\section{Parameter point selection and stability evaluation}
\label{sec:method}

We categorize the stability or metastability of a parameter point by a
 multi-stage process. First, a consistent set of Lagrangian parameters at a
 fixed renormalization scale is generated by
 \spheno~\cite{Porod:2003um, Porod:2011nf}, such that the MSSM physics at the
 DSB vacuum is consistent with the SM inputs ($m_{Z}, G_{F}$, \etc), and these
 parameters are stored in a file in the \slha format which is passed to
 \vcs, using a model file automatically generated by
 \sarah~\cite{Staub:2008uz, Staub:2009bi, Staub:2010jh, Staub:2012pb,%
 Staub:2013tta}; for consistency of input, the version of \spheno was also
 generated by \sarah. \vcs is a publicly-available
 code~\cite{Camargo-Molina:2013qva} that then prepares the minimization
 conditions for the tree-level potential as input for the publicly-available
 binary \homps~\cite{lee2008hom4ps} that finds all possible solutions to the
 particular minimization conditions of the parameter point. These are then used
 by \vcs as starting points for gradient-based minimization by
 \minuit~\cite{James:1975dr} through \pyminuit~\cite{pyminuit} to minimize the
 full one-loop potential with thermal corrections at a given temperature. If a
 minimum deeper than the DSB vacuum is found, the probability of tunneling out
 of the false DSB vacuum is then calculated through
 \ct~\cite{Wainwright:2011kj}. For a full discussion of the calculation of the
 bounce action and its conversion to a tunneling time from a false vacuum to a
 true vacuum, we refer the reader to the \vcs
 manual~\cite{Camargo-Molina:2013qva}, the \ct manual~\cite{Wainwright:2011kj},
 and the seminal papers on tunneling out of false
 vacua~\cite{Coleman:1977py, Callan:1977pt}.

If a parameter point is found to have a deeper CCB minimum, we label it as
 metastable, otherwise we label it stable\footnote{It may be that a parameter
 point is actually metastable if other scalar fields such as the partners of
 bottom quarks were allowed non-zero \vevs. However, we restrict ourselves to
 a region of parameter space where such concerns are negligible as the
 relevant trilinear interaction is small, but note also that this restriction
 cannot mistakenly label a stable parameter point as metastable.}. We then
 divide the metastable points into short-lived points which would tunnel out
 of the false DSB vacuum in three giga-years or less (corresponding to a
 survival probability of lasting $13.8$ Gy of one per-cent or less), and the
 rest as long-lived. Finally, we divide the long-lived points into thermally
 excluded, by having a probability of the DSB vacuum surviving thermal
 fluctuations of one per-cent or less, or allowed, by having a survival
 probability of greater than one per-cent, as described in more detail in the
 following subsection.

\subsection{Thermal corrections}
\label{sec:thermal_corrections}

Since the temperature of the Universe has been negligible for most of its
 existence, it is quite reasonable to calculate the tunneling time assuming
 that the four-dimensional bounce action $S_{4}$ is the dominant contribution
 to the decay width of the false vacuum. However, for sufficiently high
 temperatures, the dominant contribution may come from solitons that are $O(3)$
 cylindrical in Euclidean space rather than $O(4)$
 spherical~\cite{Linde198137}.

If the thermal contribution dominates, the expression for the decay width per
 unit volume $\Gamma / V$ at a temperature $T$ changes accordingly:
\begin{equation}
\Gamma / V = A e^{-S_{4}} \to \Gamma(T) / V(T) = A(T) e^{-S_{3}(T)/T}
\end{equation}
 where $A$ is a quantity of energy dimension four, which is related to the
 ratio of eigenfunctions of the determinants of the action's second functional
 derivative, and $S_{3}(T)$ is the bounce action integrated over three
 dimensions rather than four, with the integration over time simply replaced by
 division by temperature because of the constant value along the Euclidean time
 direction. The leading thermal corrections to the potential are at one loop,
 and given by
\begin{equation}
\Delta V(T) = \sum T^{4} J_{\pm}( m^{2} / T^{2} ) / ( 2 {\pi}^{2} )
\end{equation}
 where the sum is over degrees of freedom: bosons as sets of real scalars,
 fermions as sets of Weyl fermions, and
\begin{equation}
J_{\pm}(r) = \pm \int_{0}^{\infty} \text{d}x \; x^{2}
 \ln\left( 1 \mp e^{-\sqrt{x^{2} + r}} \right)
\end{equation}
 with $J_{+}$ for a real bosonic degree of freedom and $J_{-}$ for a Weyl
 fermion (note that we incorporate the negative sign into the definition of
 $J_{-}$ in contrast to \Ref~\cite{Brignole:1993wv}). The probability
 $P(T_{i},T_{f})$ of not tunneling between the time when the Universe is at
 temperature $T_{i}$ and when it is at temperature $T_{f} < T_{i}$ becomes
\begin{equation}
\label{prob:thtunnel}
P(T_{i},T_{f})
 = \exp\left( -\int_{T_{i}}^{T_{f}} \frac{\text{d}t}{\text{d}T}
 V(T) A(T) e^{-S_{3}(T)/T} \text{d}T \right).
\end{equation}

\subsubsection{Evaluating the survival probability}

Even the numerical evaluation of the action is computationally intense and
 while one could attempt to numerically integrate \Eq{\ref{prob:thtunnel}},
 this is impractical for more than a handful of parameter points. Hence we
 exclude parameter points based on an upper bound on the survival probability
 under some approximations, which requires $S_{3}(T)$ to be evaluated only
 once.

Firstly, the factor $A(T)$ is taken to be $T^{4}$, as the evaluation of the
 eigenfunctions of the determinant is so hard that they are usually estimated
 on dimensional grounds anyway, which is justified as the exponent of the
 action is much more important \cite{Linde:1981zj}. Any deviation would
 effectively contribute $\ln(A T^{-4})$ to $S_{3}(T)/T$, and $S_{3}(T)/T$ is
 $\sim 240$ for survival probabilities that are not extremely close to zero or
 one.

Secondly, we assume that the Universe is radiation dominated during its
 evolution from $T_i$ to $T_f$ and that entropy is approximately conserved
 between $T_i$ and today, as it is appropriate for the MSSM. Entropy
 conservation implies that $V(T_0)/V(T)=s(T)/s(T_0)$, where $s$ is the entropy
 density and $T_0=2.73\,{\rm K}$ is the temperature of the Universe today.
 Using the relation for $\text{d}t/\text{d}T$ during radiation domination, we
 can replace in \Eq{\ref{prob:thtunnel}}
\begin{equation}
\frac{\text{d}t}{\text{d}T} V(T)
 = -M_{\text{Planck}} \sqrt{90 / (  {\pi}^{2} g_{\ast}(T) )} T^{-3} V(T_0)
 \frac{s(T_0)}{s(T)}\,,
\end{equation}
where $M_{\text{Planck}}$ is the reduced Planck mass.
 The volume of the presently observable Universe (defined through
 the co-moving horizon) with 68.3\% Dark Energy 
 and 31.7\% non-relativistic matter is
 $V(T_0)=141.4 (H(T_0))^{-3}=(3.597 \times 10^{42} / \gev)^{3}$, where
 $H(T_0)=0.68 \times 100\,{\rm km}({\rm s}\, {\rm Mpc})^{-1}$, and
 the ratio $s(T_0)/s(T)$ is taken as
 $(g_{\ast s}(T_{0}) T_{0}^{3})/(g_{\ast s}(T) T^{3})$ and
 $g_{\ast s}(T_{0}) = 43/11$.
 
The tunneling is assumed to be dominated at a temperature above that at
 which the DSB vacuum evaporates, so all degrees of freedom of the SM are taken
 to be relativistic, while non-SM particles are assumed to be still
 non-relativistic at this temperature. This is because if the dimensionful
 terms such as soft SUSY-breaking terms are of the order of some scale $Q$, the
 CCB minimum should be deeper than the DSB vacuum by about $Q^{4}$ and
 effective thermal contributions to the masses of about $T$ are likely to make
 tunneling impossible by $T \simeq Q$. Hence $T$ should be less than the
 typical masses of the non-SM particles. Thus
 $g_{\ast s}(T)\equiv g_{\ast}(T)=106.75$, entirely due to the SM particles.

Putting it all together, we take
\begin{align}
\nonumber 
&\int_{T_{i}}^{T_{f}} \frac{\text{d}t}{\text{d}T}
 V(T) A(T) e^{-S_{3}(T)/T} \text{d}T \\
& \simeq 1.581 \times 10^{106} \gev
 \int_{T_{f}}^{T_{i}} T^{-2} e^{-S_{3}(T)/T} \text{d}T  .
\end{align}

Thirdly, as the evaluation of $S_{3}(T)$ is very costly in CPU time, we assume
 that $S_{3}(T)$ is a monotonically increasing function of $T$. As the
 magnitudes of the field values increase along the path from the false vacuum
 to the CCB vacuum, the masses of the degrees of freedom increase (barring
 occasional cancellations). Hence the thermal contributions lower the effective
 potential \textit{less} near the CCB vacuum than near the false vacuum, hence
 \textit{increasing} $T$ leads to the absolute height of the energy barrier
 decreasing but the barrier height relative to the false vacuum, which is the
 important quantity, \textit{increases}, and thus $S_{3}(T)$ increases.
\begin{align}
&\int_{T_{f}}^{T_{i}} T^{-2} e^{-S_{3}(T)/T} \text{d}T
  >  \int_{T_{f}}^{T_{i}} T^{-2} e^{-S_{3}(T_{i})/T} \text{d}T \nonumber \\
& \hspace{2cm}  =  ( e^{-S_{3}(T_{i})/T_{i}}
 - e^{-S_{3}(T_{i})/T_{f}} ) / S_{3}(T_{i}) \nonumber \\
 \\
& \int_{0}^{T_{i}} T^{-2} e^{-S_{3}(T)/T} \text{d}T
  >  e^{-S_{3}(T_{i})/T_{i}} / S_{3}(T_{i}) .
\end{align}

Given this,
\begin{align}
\nonumber
\label{prob:thbound}
& P(T_{i} = T,T_{f} = 0)   <  \exp\big( -1.581 \times 10^{106} \gev \\
& \hspace{4cm} \times e^{-S_{3}(T)/T} / S_{3}(T) \big) \nonumber \\
 & =  \exp\left( -\exp[ 244.53
 - S_{3}(T)/T - \ln( S_{3}(T) / \gev ) ] \right) \nonumber \\
\end{align}
 and all that remains is to find the optimal \BOL{$T = T_{\text{opt}}$} to
 maximize this quantity to find an upper bound on the survival probability
 $P(T_{i} = T_{\text{opt}},T_{f} = 0)$ for the DSB vacuum. Hence if we can
 choose $T_{\text{opt}}$ before attempting to calculate
 $S_{3}(T_{\text{opt}})$, we only need make one evaluation of
 $S_{3}(T_{\text{opt}})$.

The evaluation of the three-dimensional bounce action along a straight path in
 ``field space'' from the false vacuum to the true
 vacuum\footnote{The full set of equations of motion of the critical bubble are
 not solved by this path~\cite{Wainwright:2011kj}, but would be solved by
 adding a term to the effective potential raising the energy barrier away from
 this path in the appropriate way. A critical bubble of the unmodified
 potential must then have an action less than the action for a critical bubble
 for the modified potential.}, denoted $S_{3}^{\text{straight}}(T)$, is much
 quicker to calculate than searching for the optimal path, so for each
 parameter point $S_{3}^{\text{straight}}(T)$ was calculated for a set of
 temperatures between the temperature at which the DSB vacuum evaporates and
 the critical temperature $T_{\text{crit}}$ at which tunneling to the CCB
 minimum becomes impossible, then was fitted as $(T_{\text{crit}} - T)^{-2}$
 times a polynomial in $T$, since the action should diverge as
 $(T_{\text{crit}} - T)^{-2}$ as $T$ approaches
 $T_{\text{crit}}$~\cite{Linde:1981zj, MikkoNotes}. This fitted function
 was then numerically minimized to estimate the value of $T = T_{\text{opt}}$
 which minimizes $P(T_{i} = T_{\text{opt}},T_{f} = 0)$, which was then used to
 evaluate the right-hand side of \Eq{\ref{prob:thbound}}, taken as the upper
 bound on the survival probability of the false vacuum.

\BOL{The estimated optimal $T_{\text{opt}}$ was then used to evaluate
 $S_{3}(T_{\text{opt}})$ properly, along the correct tunneling path (not the
 straight path) between the CCB vacuum at temperature $T_{\text{opt}}$ (found
 by gradient-based minimization of the full one-loop thermal potential starting
 from the minimum at $T = 0$) and the ``DSB vacuum at $T_{\text{opt}}$'', which
 is where gradient-based minimization starting from the position of the DSB
 vacuum at $T = 0$ ends up: above the evaporation temperature, this should be
 the field origin, and indeed was for each parameter point, also demonstrating
 that the field origin is a true minimum of the potential at
 $T = T_{\text{opt}}$.}

The above procedure has been incorporated into version $1.1$ of \vcs and has
 been made public for download from HepForge.

\subsubsection{Range of validity}

As discussed in \Ref~\cite{Camargo-Molina:2013qva}, one should not trust a
 fixed-order loop expansion for \vevs very much larger than the renormalization
 scale $Q$. Likewise, thermal tunneling dominated at temperatures $T \gg Q$
 might not be very accurate. One would hope that the incorporation of running
 parameters and leading logarithmic corrections to the thermal
 contributions~\cite{Brignole:1993wv} would stabilize the results acceptably.
 While we are working on extending \vcs to include these enhancements, the
 results presented here are based purely on the one-loop effective potential
 with running parameters evaluated at a fixed $Q$. However, for every single
 one of our parameter points, the \vevs of the CCB minima were within a factor
 of a few of $Q$ and the thermal tunneling was also dominated by
 $T \lesssim Q$. \BOL{Hence the logarithms associated with higher orders are
 not large, and the one-loop expansion of the thermal potential is valid
 throughout the entire field space considered.}

We note that exclusion based on thermal tunneling is dependent on the thermal
 history of the Universe: if combined with a model where the Universe is never
 hot enough to allow tunneling at the optimal $T$, the parameter point is still
 valid. \BOL{Indeed, given appropriate initial conditions, consistency with big
 bang nucleosynthesis requires reheat temperatures only above a few MeV (see
 \eg}~\cite{Hannestad:2004px}\BOL{). The exclusions presented here are
 nonetheless important since in the most commonly hypothesized cosmologies,}
 $T \sim 10^{5} \gev$ is already considered very
 low~\cite{Falk:1996zt, Davidson:2002qv, Feng:2003zu, Pradler:2006qh,%
 Buchmuller:2007ui, Olechowski:2009bd}.

Finally, we do not address the question of whether there are additional CCB
 minima at extremely large \vevs $\gtrsim 10^{16} \gev$ which can only be
 reliably calculated with current methods using running parameters and even
 then only under restricted circumstances~\cite{Ellis:2008mc}, nor do we
 consider the effects of inflation and re-heating~\cite{Falk:1996zt}.

\subsection{Parameter scan}
\label{sec:parameter_scan}

While spontaneous symmetry breaking in the SM is triggered by a negative
 mass-squared term in the Lagrangian for the Higgs
 field\footnote{The possibility that it is due to a massless Coleman-Weingberg
 model has been ruled out by measurements of the top mass, for
 example~\cite{Sher:1988mj}.}, it is neither a necessary nor sufficient
 condition for any scalar field in a multi-scalar theory to develop a non-zero
 \vev~\cite{CamargoMolina:2012hv}. In particular, a positive mass-squared for
 the stop fields does not preclude a parameter point from having a CCB minimum,
 especially if the trilinear couplings $T_{U33} = Y_{t} A_{t}$ and $Y_{t} \mu$
 for $H_{u} \sQ \sU$ and $H_{d} \sQ \sU$ respectively are large compared to the
 square roots of the soft SUSY-breaking mass-squareds $m^2_{Q33}$ and
 $m^2_{U33}$.


Given then that we are investigating the Natural MSSM and restricting ourselves
 to the possibility of tunneling to minima with \tsq~\vevs, we choose the
 region in parameter space described by \Tab~\ref{tab:parameter_ranges}. The
 large value of the pseudoscalar Higgs mass places the scan firmly in the
 decoupling regime of the MSSM Higgs sector~\cite{Djouadi:2005gj}. To ensure
 that scalar partners other than the stops are not relevant to the analysis, we
 set them to have large masses-squared and zero soft SUSY-breaking trilinear
 interactions. Since the gluino also can have a non-negligible contribution to
 the mass of the lightest scalar Higgs, we chose to keep it at $1000 \gev$ and
 took masses for the other gauginos roughly according to a typical hierarchy
 that is expected from unification of the gauge forces~\cite{Martin:1997ns}.
 Our parameter scan thus largely overlaps with that of
 \Ref~\cite{Chowdhury:2013dka}.

\begin{table}
\centering
\begin{tabular}{
c@{\hspace{4mm}}c@{\hspace{4mm}}@{--}@{\hspace{4mm}}c}
 Parameter & \multicolumn{2}{c}{Range} \\
\hline
 $\tan\beta$ & 5 & 60 \\
\hline
 $m^{2}_{Q33}$ & $500^{2} \gev^{2}$ & $1500^{2} \gev^{2}$ \\
\hline
 $m^{2}_{U33}$ & $500^{2} \gev^{2}$ & $1500^{2} \gev^{2}$ \\
\hline
 $\mu$ & 100 \gev & 500 \gev \\ 
\hline
 $T_{U33}$ & -3000 \gev & 3000 \gev \\
\hline
\end{tabular}
\caption{Parameter ranges used in the scan. The soft SUSY-breaking mass-squared
 parameter for the $SU(2)_{L}$ doublet squarks is given by $m^{2}_{Q}$ and that
 of the $SU(2)_{L}$ singlet up-type squarks by $m^{2}_{U}$. All mass-squared
 matrices for the scalar partners of SM fermions were diagonal, and all
 diagonal entries but those shown above were set to $1500^{2} \gev^{2}$. The
 soft SUSY-breaking mass terms for the $U(1)_{Y}$, $SU(2)_{L}$, and $SU(3)_{c}$
 gauginos were $100 \gev$, $300 \gev$, and $1000 \gev$, respectively. The soft
 SUSY-breaking coefficient for the trilinear $H_{u} \sQ \sU$ interaction
 $T_{U33}$ is often written as $A_{t} \times Y_{t}$; all other soft
 SUSY-breaking trilinear terms were set to zero. Finally, the mass of the
 pseudoscalar Higgs boson was set to $1000 \gev$. The renormalization scale for
 each parameter point was the mean of the physical \tsq masses at the DSB
 vacuum.
 \label{tab:parameter_ranges}}
\end{table}

\subsubsection{Comparison in methodology to previous works}
\label{sec:comparison_to_previous}

Much early work in the area of tunneling to CCB minima in the MSSM focused on
 analytic expressions derived from the tree-level potential to determine
 whether there would be a CCB global
 minimum~\cite{AlvarezGaume:1983gj, Derendinger:1983bz, Kounnas:1983td,%
 Kounnas:1983wd, Drees:1985ie, Komatsu:1988mt}, though it has been known for
 some time that such expressions are neither necessary nor
 sufficient~\cite{Gunion:1987qv, Langacker:1994bc}, and only general outlines
 of algorithms could be given~\cite{Casas:1995pd}. It has also been known for
 some time that they gave no hint as to whether the tunneling time out of the
 DSB false vacuum could be phenomenologically
 acceptable~\cite{Claudson:1983et, Kusenko:1996jn}. 

The algorithm used by \vcs improves upon these by finding \textit{all} the
 minima of the tree-level potential, not just those that may lie on special
 lines in field space, as well as incorporating loop corrections, which,
 despite various claims in the
 literature~\cite{PhysRevD.48.4352, Casas:1995pd}, are
 important~\cite{Bordner:1995fh, Ferreira:2000hg, Camargo-Molina:2013sta}.

One may note the overlap in objective with the works of
 \Refs~\cite{Blinov:2013fta} and~\cite{Chowdhury:2013dka}: CCB minima with stop
 \vevs are searched for in a similar parameter space, and metastable points are
 categorized as acceptably long-lived or not based on tunneling times
 calculated by \ct. The major improvement over these works is that we also
 exclude points based on low probabilities to survive thermal fluctuations when
 the Universe was at a temperature of the order of $1$ TeV. However, we also
 note that we improve upon the zero-temperature results of these works in two
 significant ways: the first is that our use of the homotopy continuation
 method guarantees that we find all the minima of the tree-level potential, as
 opposed to a random seeding of the field space followed by gradient
 minimization in \Ref~\cite{Blinov:2013fta}, which obviously cannot guarantee
 that the random seeding did not miss a CCB minimum, or a brute-force
 four-dimensional grid scan in \Ref~\cite{Chowdhury:2013dka}, which may miss
 minima just beyond the range of the grid. The second way is that we use the
 full one-loop effective potential rather than the tree-level potential. Though
 one would hope that the loop corrections do not significantly alter the
 tree-level conclusions, it is not always the case, and the tree-level results
 can be rather sensitive to the renormalization scale chosen for the running
 parameters, while the loop corrections stabilize the dependence on the
 scale~\cite{Camargo-Molina:2013sta}.

%% file: results.tex
 \begin{figure*}[t]
\centering
\includegraphics[width=0.49\linewidth
]{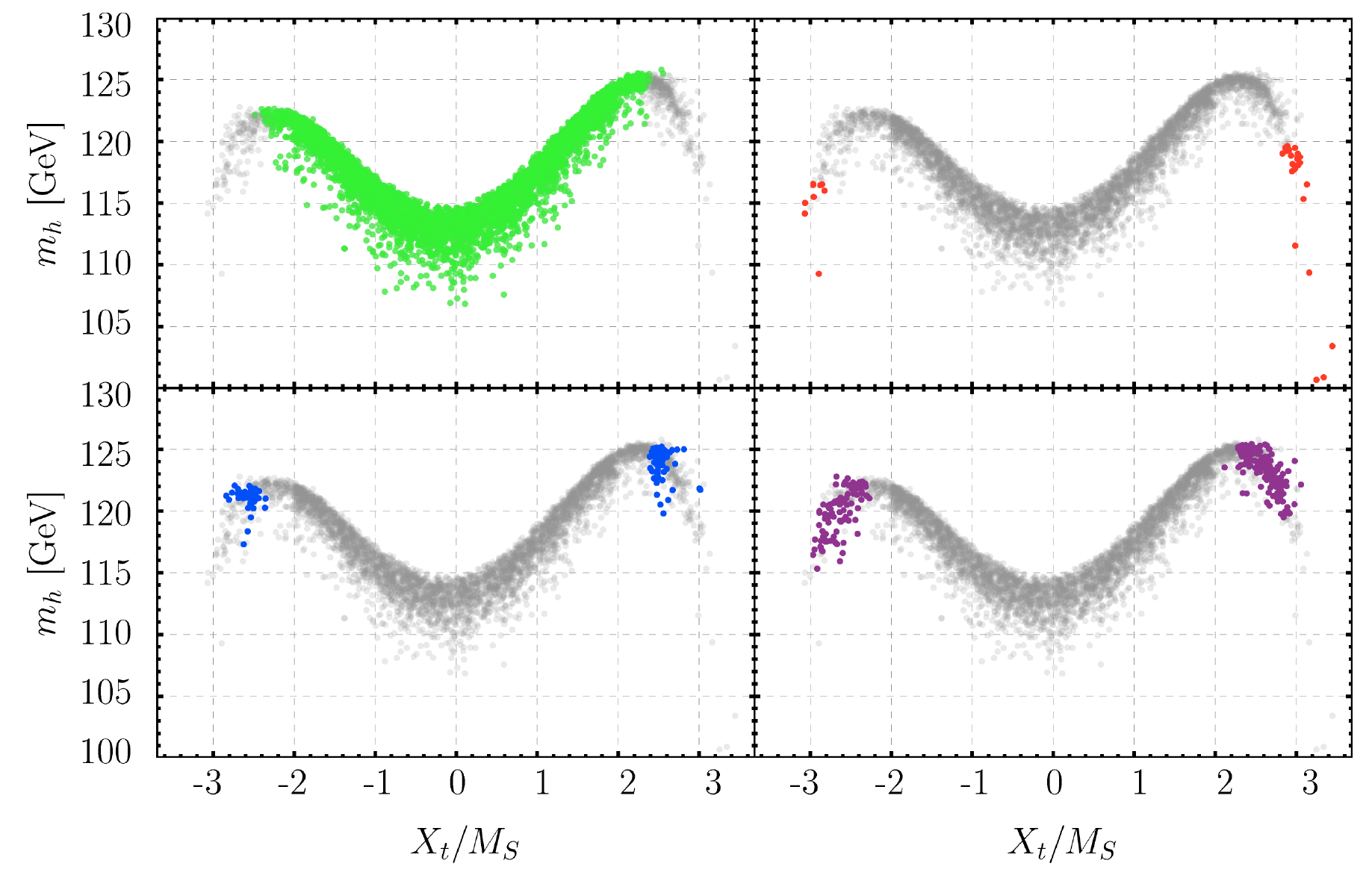} \hfill
\includegraphics[width=0.49\linewidth
]{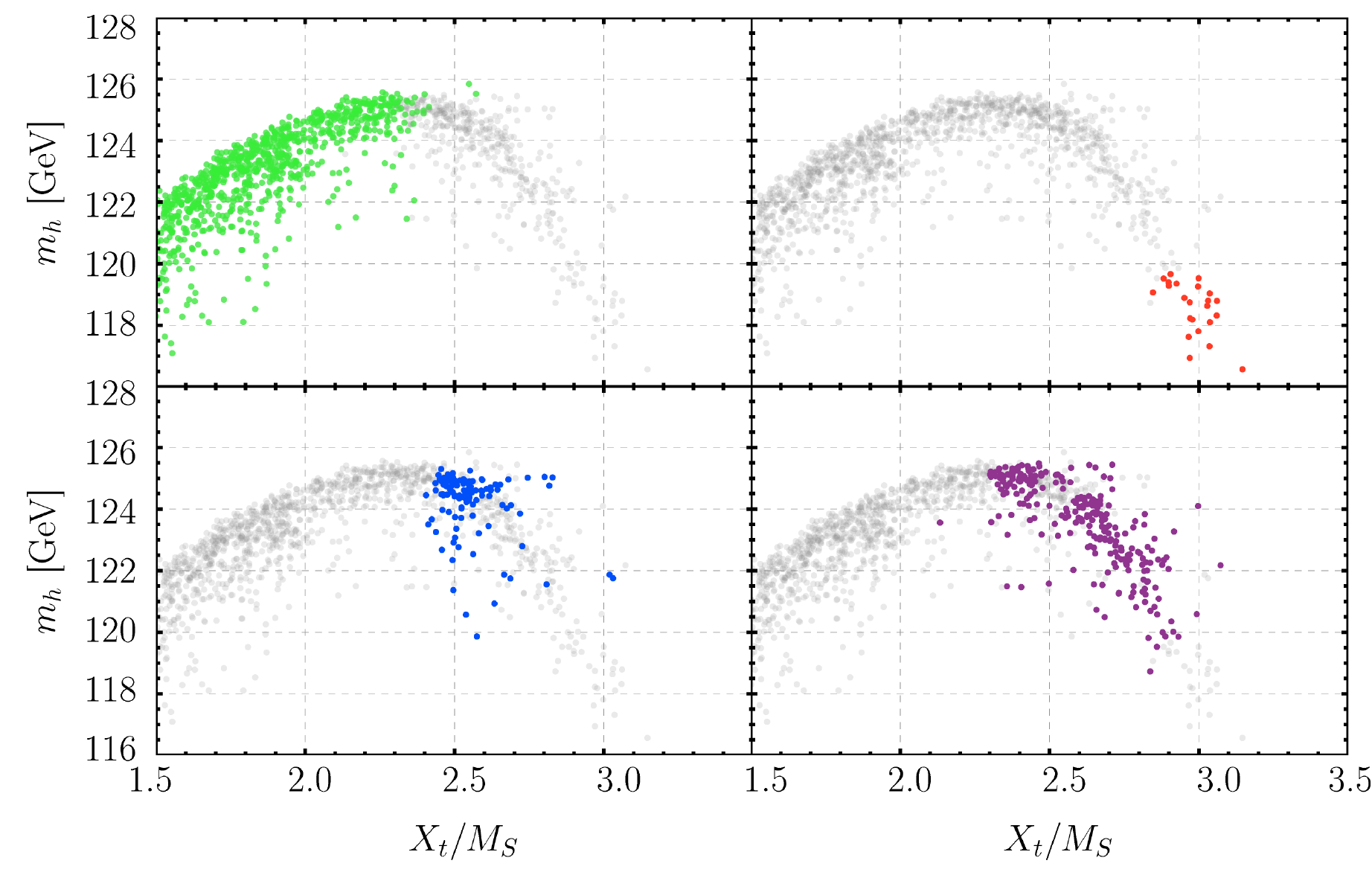}
\caption{Categorization of parameter points as to whether they are allowed or
 excluded by tunneling out of the DSB vacuum.
 \BOL{Green (top left): no CCB minimum deeper than the DSB minimum was found.
 Blue (bottom left): the DSB minimum is a false vacuum, but the
 probability of surviving 13.8 Gy at zero temperature and surviving thermal
 fluctuations are both above one per-cent. 
 Purple (bottom right): the probability of surviving tunneling out of the DSB
 false vacuum at non-zero temperature is less than one per-cent. 
 Red (top right): the probability of the DSB false vacuum surviving $13.8$ Gy
 at zero temperature is less than one per-cent. On the right we zoom in on the
 region with $X_t/M_S \in [1.5,3.5]$  and $m_h \in [116, 128] \gev$.}
 \label{fig:Xt_vs_mh_rpbg}}
\end{figure*}

\section{Constraining the parameter space of the Natural MSSM}
\label{sec:results}

Our primary result is that a large proportion of the parameter space where the
 Higgs boson mass is even slightly compatible with the measurement of
 $125 \gev$~\cite{Aad:2012tfa,Chatrchyan:2012ufa} is ruled out by thermal
 tunneling even though the tunneling time at zero temperature is much longer
 than the observed age of the Universe. This is presented in
 \Fig~\ref{fig:Xt_vs_mh_rpbg}, where the parameter points of our
 five-dimensional scan are projected onto a two-dimensional plane with the axes
 being the mass of the lightest Higgs scalar $m_{h}$ and the ratio
 $X_{t} / M_{S}$, where $X_{t} = A_{t} - \mu \cot\beta$, and $M_{S}$ is the
 square root of the product of the tree-level \tsq masses evaluated at the DSB
 minimum as this should keep higher order corrections
 small~\cite{Gamberini:1989jw}. One would expect this ratio to be correlated
 with the probability of tunneling out of the DSB vacuum, as a combination of
 the comparisons mentioned in \Sec~\ref{sec:parameter_scan}. (Though tunneling
 was evaluated at one loop, the value of $m_{h}$ for each parameter point is
 based on a full diagrammatic one-loop calculation including the effects of the
 external momenta~\cite{Pierce:1996zz} and, in addition, the known two-loop
 corrections are
 included~\cite{Degrassi:2001yf, Brignole:2002bz, Brignole:2001jy,%
 Dedes:2002dy}.)


However, while increasing $|X_{t}| / M_{S}$ is correlated with decreasing
 stability in some sense, in the phenomenologically interesting region where
 $m_{h} > 123 \gev$, it fails to discriminate \BOL{effectively} between
 acceptable points with stable or high survival probability DSB vacua and those
 with low survival probabilities for their DSB vacua. \BOL{In fact, no
 projection of our scan onto a plane in terms of simple combinations of the
 input parameters showed any clear discriminatory power, and thus we conclude
 that a full calculation is inevitably necessary.}

\subsection{Comparison to previous results}

\begin{figure}[tbp]
\centering
\includegraphics[width=0.7\linewidth
]{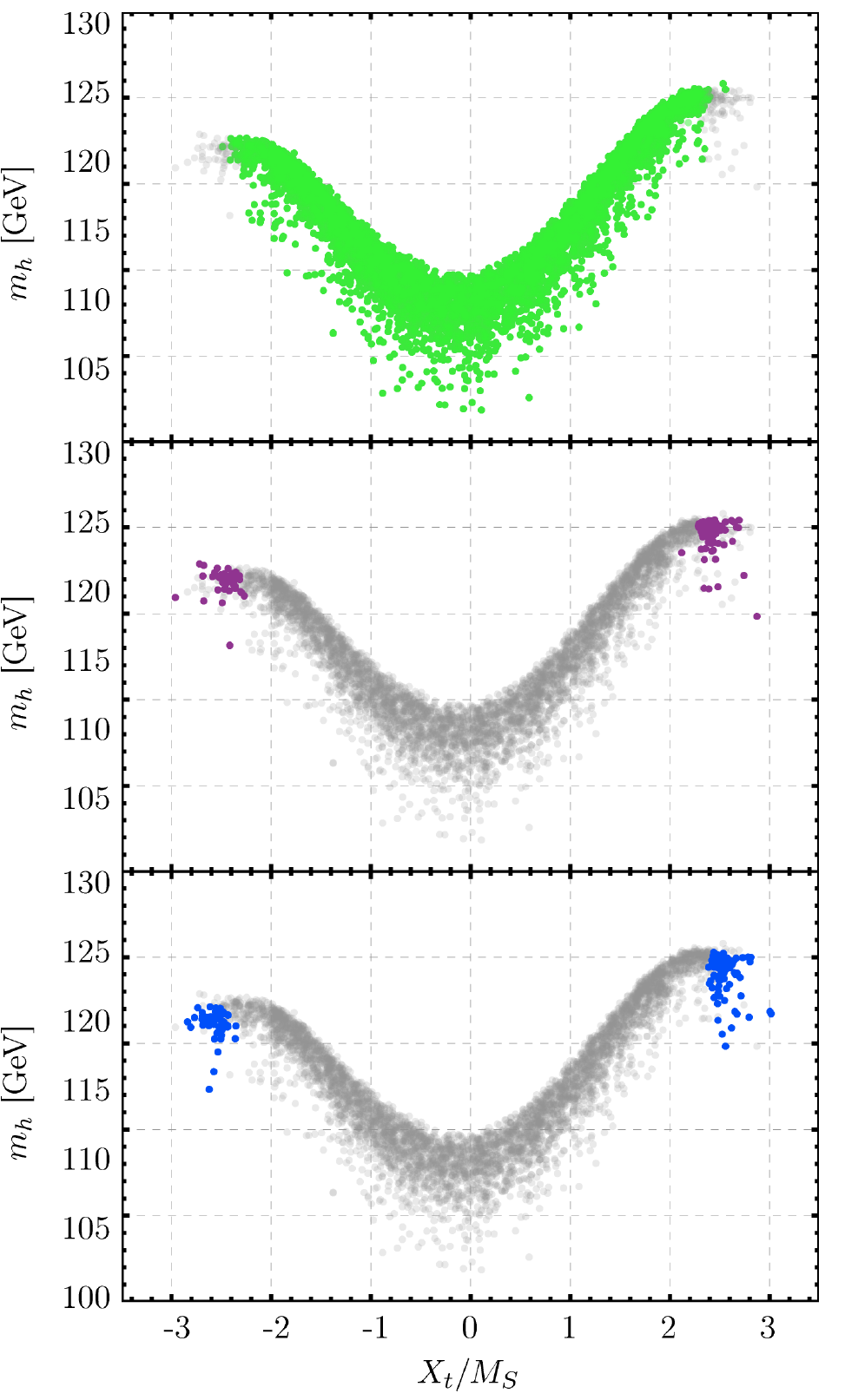}
\caption{\BOL{Results without displaying those points
 which would be excluded by condition (\ref{eq:analytic_thumb_rule}).
 The color coding is the same as in \Fig~\ref{fig:Xt_vs_mh_rpbg},
 with green points contrasted against the others in the top picture,
 purple in the middle, and blue in the bottom picture. We note that this
 condition excludes all points with short-lived DSB vacua at zero temperature.}
 \label{fig:Xt_vs_mh_after_thumb_rule}}
\end{figure}

Even though it was derived under the assumption that the Yukawa coupling is
 much smaller than the gauge couplings, which is obviously wrong for the top
 sector, and even though it has been known to be neither necessary nor
 sufficient~\cite{Gunion:1987qv, Langacker:1994bc}, the condition
\begin{equation}
\label{eq:analytic_thumb_rule}
 A_{t}^{2} < 3 ( m^{2}_{Q33} + m^{2}_{U33} + m^{2}_{H_{u}} )
\end{equation}
 has been used in place of a proper analysis as a check that parameter points
 have stable DSB vacua. It has been demonstrated numerically that it is neither
 necessary nor sufficient, nor meaningfully correlated with long-/short-lived
 metastable vacua~\cite{Camargo-Molina:2013sta}, but for completeness we show
 how our results are if we exclude points which fail the condition in
 \Fig~\ref{fig:Xt_vs_mh_after_thumb_rule}. Coincidentally, the condition
 happens to exclude all the points with DSB vacua that are short-lived at zero
 temperature, but it both unnecessarily excludes stable and acceptably
 long-lived metastable points at larger $|X_{t}|$ and fails to exclude most of
 the points which are excluded by thermal tunneling with $m_{h} > 123 \gev$.


An attempt to account for acceptably long-lived DSB vacua by empirically
 fitting coefficients~\cite{Kusenko:1996jn} led to the following condition:
\begin{equation}
\label{eq:empirical_thumb_rule}
 A_{t}^{2} + 3 {\mu}^{2} < 7.5 ( m^{2}_{Q33} + m^{2}_{U33} )
\end{equation}
 but not even one of the points in our scan was excluded by this condition,
 hence we consider it irrelevant. \BOL{Hence we stress again that one
 should not rely on analytic conditions which are derived using simplifying
 assumptions or which are based on ostensible patterns found in a particular
 numerical analysis. For a serious check of the stability of the scalar
 potential, a full-fledged numerical evaluation for each point is usually
 inevitable. However, while the typical running time per metastable
 point (those plotted in red, purple, or blue) per CPU core with \vcs 1.1 was
 10-30 minutes, one can easily change a setting so that it will evaluate
 whether a parameter point is stable or metastable (green or not) within
 seconds, which for example should be sufficient for purposes of finding
 conservatively acceptable parameter regions.}

If we ignore thermal tunneling, our results qualitatively agree with
 \Refs~\cite{Blinov:2013fta} and~\cite{Chowdhury:2013dka}. Though the parameter
 space overlap with \Ref~\cite{Blinov:2013fta} is not as great, we largely
 agree with the ratios of $X_{t}$ to $M_{S}$ where the CCB minima become deeper
 than the DSB minima and where the tunneling time becomes unacceptably short.
 Likewise, we agree with the ratios one can read off the figures in
 \Ref~\cite{Chowdhury:2013dka}, but note that the values of $m_{h}$ therein
  are inconsistent with our calculation (using
 \spheno), the calculation in \Ref~\cite{Blinov:2013fta} (using \suspect), and
 the results in \Ref~\cite{Arbey:2011ab} (using \suspect and
 \softsusy)\footnote{The mismatch in $m_{h}$ is under investigation by the
 authors of \Ref~\cite{Chowdhury:2013dka} and \suseflav~\cite{private}.}.
 \BOL{The Higgs masses calculated by \softsusy, \spheno, and \suspect are
 within the theoretical uncertainity of 2--4 \gev using two-loop 
 corrections. In contrast, the differences between these codes and
 {\tt FeynHiggs}~\cite{Heinemeyer:1998yj, Hahn:2009zz} are usually
 larger because of the different renormalization scheme: the difference for
 \Fig~\ref{fig:Xt_vs_mh_rpbg} is a steady increase in $m_h$ with increasing
 $|X_{t}/M_{S}|$, with good agreement for $|X_{t}/M_{S}| = 0$, to an increase
 of about $3 \gev$ for $X_{t}/M_{S} = -2.5$ and an increase of about $5 \gev$
 for $X_{t}/M_{S} = +2.5$.}

%% file: discussion.tex
\section{Conclusion}
\label{sec:discussion}

We have presented an exploration of what regions of the Natural MSSM parameter
 space can be excluded by demanding at least a one per-cent survival
 probability for the vacuum with the desired symmetry breaking against
 tunneling to charge- and color-breaking vacua at non-zero temperatures. In
 order to do so, we extended the feature set of \vcs to include the
 functionality to exclude parameter points based on thermal tunneling, which we
 have made publicly available: \vcs $1.1$ is available for download from
 HepForge:
\begin{center}
 {\tt http://www.hepforge.org/downloads/vevacious}.
\end{center}

Stability against thermal tunneling is a relevant constraint, especially in the
 parameter space of the MSSM where the mass of the lightest Higgs boson is
 consistent with observations. While exclusion based on zero-temperature
 tunneling can also exclude regions of the parameter space, points that have
 sufficiently long lifetimes at zero temperature may have very low probability
 to avoid ending up in a CCB vacuum by the time the temperature drops to a
 negligible value. Unfortunately, the dependence on the Lagrangian parameters
 is not simple, and a full analysis of any given parameter point seems
 necessary, though straightforward given the availability of \vcs.

We have also showed that results on metastability based on previous tree-level
 analyses are not significantly affected by the zero-temperature one-loop
 corrections, as opposed to the effects at finite-temperature.